\documentclass[apj]{emulateapj}
\def\ltsima{$\; \buildrel < \over \sim\;$}
\def\ltsim{\lower.5ex\hbox{\ltsima}}
\def\gtsima{$\; \buildrel > \over\sim \;$}
\def\gtsim{\lower.5ex\hbox{\gtsima}}
\usepackage{amsmath}
 
\shorttitle{RELATIVISTIC FLOWS}
\shortauthors{Kikuchi and Shigeyama}

\begin{document}

\title{RELATIVISTIC FLOWS 
	AFTER SHOCK EMERGENCE }

\author{Reina Kikuchi$^{1,2}$}
\author{ Toshikazu Shigeyama$^1$}
\affil{$^1$Research Center for the Early Universe, Graduate School of Science, $^2$Department of Astronomy, Graduate School of Science,
        University of Tokyo, Bunkyo-ku, Tokyo 113-0033, Japan}

\begin{abstract}
We investigate relativistic flows after a shock wave generated in a star arrives at the surface. 
First, the sphericity effect is involved through a successive approximation procedure by adding correction terms to an already known self-similar solution in the ultra-relativistic limit assuming the plane parallel geometry. 
It is found that the lowest order correction terms are of the order of $(a/R_*)^{1/(6\sqrt{3}-8)}$ for a star with a radiative envelope, where $a$ is the distance from the stellar surface at the moment of shock breakout and $R_*$ the stellar radius. 
We also find that the involvement of sphericity increases the acceleration in the early phase as compared with that of the original plane-parallel flow. 
Second, we obtain semi-analytic solutions for a mildly relativistic flow in which the rest mass energy density is not negligible in the equation of state. To take into account this, we use the enthalpy and the pressure instead of using the density and the pressure as thermodynamic variables. 
These solutions assume self-similar evolutions except for the initial conditions. 
Third, we have carried out numerical calculations with a special-relativistic hydrodynamical code based on the Godunov method in order to check the applicability of the above sphericity corrections and the semi-analytic solutions. 
The equation of state used in our calculations includes the rest mass energy density. Comparisons with results of numerical calculations support the validity of the sphericity correction terms.  
The evolutions of the pressure and the Lorentz factor of each fluid element of the semi-analytic solution for mildly relativistic flows match the numerical results at least in early phases. 
We also investigate the final free expansion phases by this code. For either spherical symmetric or plane-parallel ultra-relativistic flows, we could not observe the free expansion phase till the self-similar variables become as large as $10^6$. 
On the contrary, the flow in which the ratio of the pressure to density at the shock breakout is less than unity reaches the free expansion. 
We have derived the final energy distributions for these flows and compare them with previous works.
\end{abstract}


\section{INTRODUCTION}
Plenty of studies have been made about shock propagation in stellar explosion. 
After the energy is liberated inside a star, a shock wave  is formed to go through toward the surface. 
When it reaches there, it disappears. 
This event is called as shock breakout or shock emergence. Material of the stellar envelope accelerated by the shock wave  expands into the surrounding space as a rarefaction wave. 
The ejected matters continue to accelerate until their thermal energy is sufficiently converted into the kinetic energy. 
Finally, they expand freely. 

First, \cite{Gandel'man56} posed this problem. 
\cite{Sakurai60} solved it with a self-similar analysis, assuming one dimensional plane-parallel (PP) configuration. 
The stellar surface marks the boundary, on one side of it the atmosphere fills and on the other side the vacuum. 
The initial density distribution is assumed to follow the power law $\rho_0\propto x^{\delta}$, where $x$ is the distance from the surface.  
Especially, $\delta=3$ represents the radiative envelope and $\delta=1.5$ convective. 
His self-similar solution is strictly restricted in applying to a real star, because sphericity is no longer ignorable in describing the flow far from the surface. 
\cite{Kazhdan92} worked to involve sphericity by adding correction terms proportionate to the power of non-dimensional parameter $x_\circ=1-r/R_*$, where $r$ is the distance from the center and $R_*$ is the stellar radius. 
\cite{Matzner99} conducted a more detailed study. 
They investigated the progenitor dependence and the sphericity involvement and derived semi-analytic distributions of physical quantities such as the velocity and the density. 
These studies ignore effects of the special relativity and the radiative cooling, which causes the speed of the front in these studies to be developed to infinity.

\cite{Tan00} studied trans-relativistic flows. 
They estimated the maximum velocity of ejecta including the effects of sphericity and radiative diffusion. 
They also studied effects of aspherical structure and gravity.

In the ultra-relativistic limit, \cite{NS05} obtained a self-similar solution ignoring the rest mass energy.   
They assumed the same initial configuration as in \cite{Sakurai60}. 
In addition, the equation of state (EOS) was assumed as $\varepsilon =3p$ in stead of $\varepsilon=3p+\rho$, where $p$ is the pressure, $\rho$ the density, and $\varepsilon$ the internal energy density per unit volume. 
This is the indispensable assumption to derive a self-similar solution. 
\cite{Sari06} presented an analytic expression for the flow after shock emergence. 

In this paper, we concentrate on the following two issues inherent to the above ultra-relativistic self-similar solutions. 
First, the assumption of spherical symmetry (SS) is a better approximation of a real system than that of PP.  
The farther a wave is from the stellar surface, the larger the deviation from the PP. 
The end of acceleration can not be described within the range that the PP approximation is valid, i.e. $|x_\circ| \ll 1$. 
The pressure varying as $p \propto \rho^{4/3} \propto (1-x_\circ)^{-4}$ does not significantly drop in that range. 
To take into account the sphericity, we need to introduce a characteristic length, the stellar radius $R_*$, which inhibits analyses by self-similar solutions.
Thus, in \S\ref{S-involve}, we conduct a successive approximation in terms of a parameter $\xi=a/R_*$ ( where $a$ is the distance of each fluid element from the stellar surface at shock emergence ) by which the sphericity is characterized. 
From the set of fundamental equations for ultra-relativistic hydrodynamics, we found that the sphericity is involved from the order 
$\cal{O}$$(\xi^{1/((2\sqrt{3}-3)\delta+1)})$ after shock emergence and $\cal{O}$$(\xi)$ before shock emergence.  
We take this effect into account only up to the order 
$\cal{O}$$(\xi^{1/((2\sqrt{3}-3)\delta+1)})$. 

Second, the acceleration does not terminate in \cite{NS05}'s solution. 
The Lorentz factor $\gamma$ of a fluid element increases infinitely with increasing time $t$ measured from the moment of shock emergence;  
$\gamma\propto t^{0.732}$ for any $\delta$. 
The source energy of acceleration is the internal energy which is converted to the kinetic energy during expansion. 
It is reasonable to consider that when the internal energy is sufficiently consumed, the ejecta are no longer accelerated. 
This implies that their self-similar solutions can not describe a stage of free expansion. 
It might result from the EOS $\varepsilon=3p$ neglecting the rest mass energy. 
The EOS has to be $\varepsilon =3p+\rho$ in later phases when the gas becomes mildly relativistic.  
In \S\ref{MR}, we derive the semi-analytic solution including this term in the EOS . 

In \S\ref{NC}, we conduct numerical calculations using the Godunov method for relativistic hydrodynamics.   
The aim is to support the validity of the sphericity involvement and the semi-analytic solutions for mildly relativistic flows. 
Calculations are carried out from the moment of shock emergence for both PP and SS configuration and  various initial energy strengths, setting $\delta=3$ and the initial conditions from the analytic solutions at $t=0$ of \cite{NS05}. 
The results are presented in \S\ref{NCresults}. 

\section{SPHERICITY INVOLVEMENT} 
\label{S-involve}
\subsection{Formulation}
In this section, we derive sphericity corrections to the PP flow  after shock emergence. 
First, we present the basic equations for an ultra-relativistic SS flow. 
Second, we transform these equations into non-dimensional ones. Third, we conduct a successive approximation to obtain 0th and 1st order equations. 
The 0th order equations provide the self-similar solutions obtained by \cite{NS05}.
The 1st ones shall provide the amount of corrections.

We confine ourselves to dealing with a perfect fluid with no external force including gravity. 
If a shock wave is strong enough, gravity has practically no effect on the fluid motion. 
We assume the initial density distribution $\rho_0=b(R_*-r)^{\delta}$. 
The basic equations of relativistic hydrodynamics  are derived from the energy-momentum and mass conservation. 
Here we present the equations for spherically symmetric flows in a fixed frame $(r,t)$.
\begin{equation}
	\frac{\partial}{\partial t}
	\left[\gamma^2(\varepsilon+\beta^2p\right)]
	+\frac{1}{r^2}\frac{\partial}{\partial r}
	\left[r^2\gamma^2(\varepsilon +p)\beta\right]=0\label{beq1},
\end{equation}

\begin{equation}
	\frac{\partial}{\partial t}\left[\gamma^2(\varepsilon +p)\beta\right]
	+\frac{1}{r^2}\frac{\partial}{\partial r}\left[
	r^2\gamma^2(\varepsilon +p)\beta^2\right]
	+\frac{\partial p}{\partial r} =0\label{beq2},
\end{equation}

\begin{equation}
	\frac{\partial \rho'}{\partial t}+\frac{1}{r^2}\frac{\partial}{\partial r}
	(r^2\rho'\beta)=0\label{beq3},
\end{equation}
where $\varepsilon$ denotes the internal energy density, $p$ the pressure
and $\rho$ the density, all measured in the fluid frame.
$\rho'=\gamma\rho$ is the density measured in the fixed frame where the fluid moves with the velocity $c\beta$. 
$\gamma$ denotes the Lorentz factor defined as $\gamma=1/\sqrt{1-\beta^2}$. 
We set the speed of light $c$ to 1 in the following.
We assume the ultra-relativistic EOS as
\begin{equation}
	\varepsilon = 3p\label{EOS}. 
\end{equation}
With this EOS (\ref{EOS}), equations (\ref{beq1}), (\ref{beq2}), and (\ref{beq3}) are transformed into the expressions presented in \cite{Blandford76};
\begin{equation}
	\frac{d}{dt}\left(p\gamma^4\right) = \gamma^2\frac{\partial p}{\partial t},
	\label{deq1}
\end{equation}

\begin{equation}
	\frac{d}{dt}\ln(p^3\gamma^4) = - \frac{4}{r^2}\frac{\partial}{\partial r}(r^2\beta),\label{deq2}
\end{equation}

\begin{equation}
	\frac{d}{dt}(p\rho^{-4/3}) = 0\label{deq3}, 
\end{equation}
where the convective derivative
\begin{equation}
	\frac{d}{dt}=\frac{\partial}{\partial t}+\beta\frac{\partial}{\partial r},
\end{equation}
has been introduced. 

In order to describe the motion after shock emergence,
it is convenient to use a Lagrangian coordinate $a$ which is the position of a fluid element at $t=0$ measured from the stellar surface.
Then from the definition of the velocity of the fluid element, we obtain
\begin{equation}
	\frac{dr}{dt}=v.\label{deq4}
\end{equation}

For a PP flow, only one non-dimensional variable $\eta$ is sufficient to describe the self-similar motion of the gas. 
In contrast, for an SS flow,  when we deal with the wave across the stellar surface, the stellar radius $R_*$ appears as a scale. The flow is not self-similar anymore. 
Another non-dimensional variable is necessary besides $\eta$. 
We can write the non-dimensional variables such as
\begin{equation}
	\eta = \frac{t}{\gamma^2_1 a},
\end{equation}
\begin{equation}
	\xi = \frac{a}{R_*}.
\end{equation}
The dependent variables are written using non-dimensional functions, $F$, $G$, $H$ and $I$, such as
\begin{equation}
	p = p_1(a)F(\eta,\xi),\label{solution1}
\end{equation}
\begin{equation}
	\gamma^2 = \gamma^2_1(a)G(\eta,\xi),\label{solution2}
\end{equation}
\begin{equation}
	\rho= \rho_1(a)H(\eta,\xi),\label{solution3}
\end{equation}
\begin{equation}
	r = (R_{\ast}-a)I(\eta,\xi).\label{solution4}
\end{equation}
Here $p_1(a)$, $\gamma^2_1(a)$ and $\rho_1(a)$ are the profiles at $t=0$.
They conform to the power law,
\begin{equation}
	p_1(a)=c_p(A,b) a^{(\delta-m)/(m+1)},
\end{equation}
\begin{equation}
	\gamma_1^2(a)=c_{\gamma}(A,b) a^{-m/(m+1)},
\end{equation}
\begin{equation}\label{density@sb}
	\rho'_1(a)=\gamma_1\rho_1=c_{\rho'}(A,b) a^{(\delta-m)/(m+1)},
\end{equation}
where
\begin{equation}
	m = (2\sqrt{3}-3)\delta,
\end{equation}
and $c_p(A,b)$, $c_{\gamma}(A,b)$ and $c_{\rho'}(A,b)$ are constants depending on $A$ and $b$. 
These two quantities characterize the flow before shock emergence. 
The parameter $A$  specifies the amount of energy of the flow, through a relation with the shock Lorentz factor $\Gamma$ by $\Gamma^2= A(-t)^m$. 
For more detailed descriptions, see \cite{NS05}. 
We transform the independent variables from $(t, a)$ to $(\eta, \xi)$ using the following relations,

\begin{equation}
	\frac{d}{dt} =\frac{\partial}{\partial \eta}\frac{\partial \eta}{\partial t}
		+\frac{\partial}{\partial \xi}\frac{\partial \xi}{\partial t}
			= \frac{\eta}{t}\frac{\partial}{\partial \eta},
\end{equation}

\begin{equation}
	\frac{\partial}{\partial a} = \frac{\partial}{\partial \eta}\frac{\partial \eta}{\partial a}
		+\frac{\partial}{\partial \xi}\frac{\partial \xi}{\partial a} 
		= -\frac{1}{m+1}\frac{\eta}{a}\frac{\partial}{\partial \eta}+\frac{\xi}{a}\frac{\partial}{\partial \xi}.
\end{equation}
Then equations (\ref{deq1}), (\ref{deq2}), (\ref{deq3}) and (\ref{deq4})
are converted to the following non-dimensional equations.

\begin{multline}
	\frac{\partial}{\partial\eta}\left(\ln F+2\ln G\right)=\\
	\frac{I^2H}{\sqrt{G}}
	\left(\frac{\delta-m}{m+1}
		-\frac{\eta}{m+1}\frac{\partial F}{\partial\eta}
	+\xi\frac{\partial F}{\partial\xi}\right),\label{ndeq1}
\end{multline}

\begin{multline}
	\frac{\partial\ln G}{\partial\eta}+\frac{3}{2}\frac{\partial\ln F}{\partial\eta}=\\
	I^2H\sqrt{G}\left(-\frac{\eta}{m+1}\frac{\partial\ln G}{\partial\eta}
		+\xi\frac{\partial\ln G}{\partial\xi}-\frac{m}{m+1}\right)\\
		-\frac{4\gamma_1^2\xi}{1-\xi}\frac{G}{I},\label{ndeq2}
\end{multline}

\begin{equation}
	\frac{\partial\ln F}{\partial\eta}
		=\frac{4}{3}\frac{\partial\ln H}{\partial\eta},\label{ndeq3}
\end{equation}

\begin{equation}
	\frac{\partial I}{\partial\eta}=\frac{\gamma_1^2\xi}{1-\xi}.\label{ndeq4}
\end{equation}
The expansions in $\gamma^{-2}$ can be truncated at the first contributing term because the motion is assumed to be ultra-relativistic. 
The last term of (\ref{ndeq2}) is of the order $\cal{O}$$(\xi^{1/(m+1)})$.
Therefore we expand the solutions (\ref{solution1}) - (\ref{solution4}) in a power series of $\xi$ starting with the term proportional to $\xi^{1/(m+1)}$.
\begin{equation}
	F(\eta,\xi)=F_0(\eta)(1+\xi^{1/(m+1)}F_1(\eta)),\label{solF}
\end{equation}
\begin{equation}
	G(\eta,\xi)=G_0(\eta)(1+\xi^{1/(m+1)}G_1(\eta)),
\end{equation}
\begin{equation}
	H(\eta,\xi)=H_0(\eta)(1+\xi^{1/(m+1)}H_1(\eta)),\label{solH}
\end{equation}
\begin{equation}
	I(\eta,\xi)=I_0(\eta)(1+\xi^{1/(m+1)}I_1(\eta)).
\end{equation}
The 0th order equations become 
\begin{equation}
	\frac{d I_0}{d\eta}=0\label{0eq1},
\end{equation}

\begin{multline}
	\frac{d\ln F_0}{d\eta}+2\frac{d\ln G_0}{d\eta}
	=\\ \frac{I_0^2H_0}{\sqrt{G_0}}
	\left(\frac{\delta-m}{m+1}-\frac{\eta}{m+1}\frac{d F_0}{d\eta}
	\right),
\end{multline}

\begin{multline}
	\frac{d\ln G_0}{d\eta}
	+\frac{3}{2}\frac{d\ln F_0}{d\eta}
	=\\ -\frac{I_0^2H_0\sqrt{G_0}}{m+1}
	\left(\eta\frac{d\ln G_0}{d\eta}-m\right),
\end{multline}

\begin{equation}
	\frac{d \ln F_0}{d\eta} 
	= \frac{4}{3}\frac{d\ln H_0}{d\eta}.
\end{equation}
These correspond to the planar ones.
The 1st order are
\begin{equation}
	\frac{d I_1}{d \eta}
	=c_{\gamma} R_*^{-m/(m+1)},
\end{equation}

\begin{multline}
	\frac{(m+1)\sqrt{G_0}}{I_0^2H_0}\left(
	\frac{d F_1}{d\eta}+2\frac{d G_1}{d\eta}
	\right)= \\	
	(2I_1+H_1-\frac{1}{2}G_1)  	
	\left(\delta-m-\eta\frac{d F_0}{d\eta}\right)  
	\\- \eta F_1\frac{d F_0}{d\eta} -\eta F_0\frac{d F_1}{d\eta} +F_0F_1,
\end{multline}

\begin{multline}
	\frac{m+1}{I_0^2H_0\sqrt{G_0}}
	\left(\frac{d G_1}{d\eta}
	+\frac{3}{2}\frac{d F_1}{d\eta}
	+4c_{\gamma} R_*^{-m/(m+1)}G_0\right)\\ =
	-\eta\frac{d G_1}{d\eta}+G_1
	\\-(2I_1+H_1+\frac{1}{2}G_1)
	\left(\eta\frac{d\ln G_0}{d\eta}+m \right),
\end{multline}

\begin{equation}
	\frac{d F_1}{d\eta}=
	\frac{4}{3}\frac{d H_1}{d\eta}.\label{1eq4}
\end{equation}
The solutions of (\ref{0eq1})-(\ref{1eq4}) are obtained through numerical integration with the following initial conditions; 
\begin{eqnarray}
	F_0(0)=G_0(0)=H_0(0)=I_0(0)=1,	\\
	F_1(0)=G_1(0)=H_1(0)=I_1(0)=0.
\end{eqnarray}
This implies that we adopt the distributions of PP geometry at $t=0$ as the initial condition of SS geometry, because we can neglect deviations from PP flows at shock emergence.
All sphericity corrections before shock emergence are $\cal{O}(\xi)$ that is higher than that after shock emergence, $\cal{O}$$(\xi^{1/(m+1)})$. 

\begin{figure}
\begin{center}
\includegraphics[clip=true,scale=0.75]{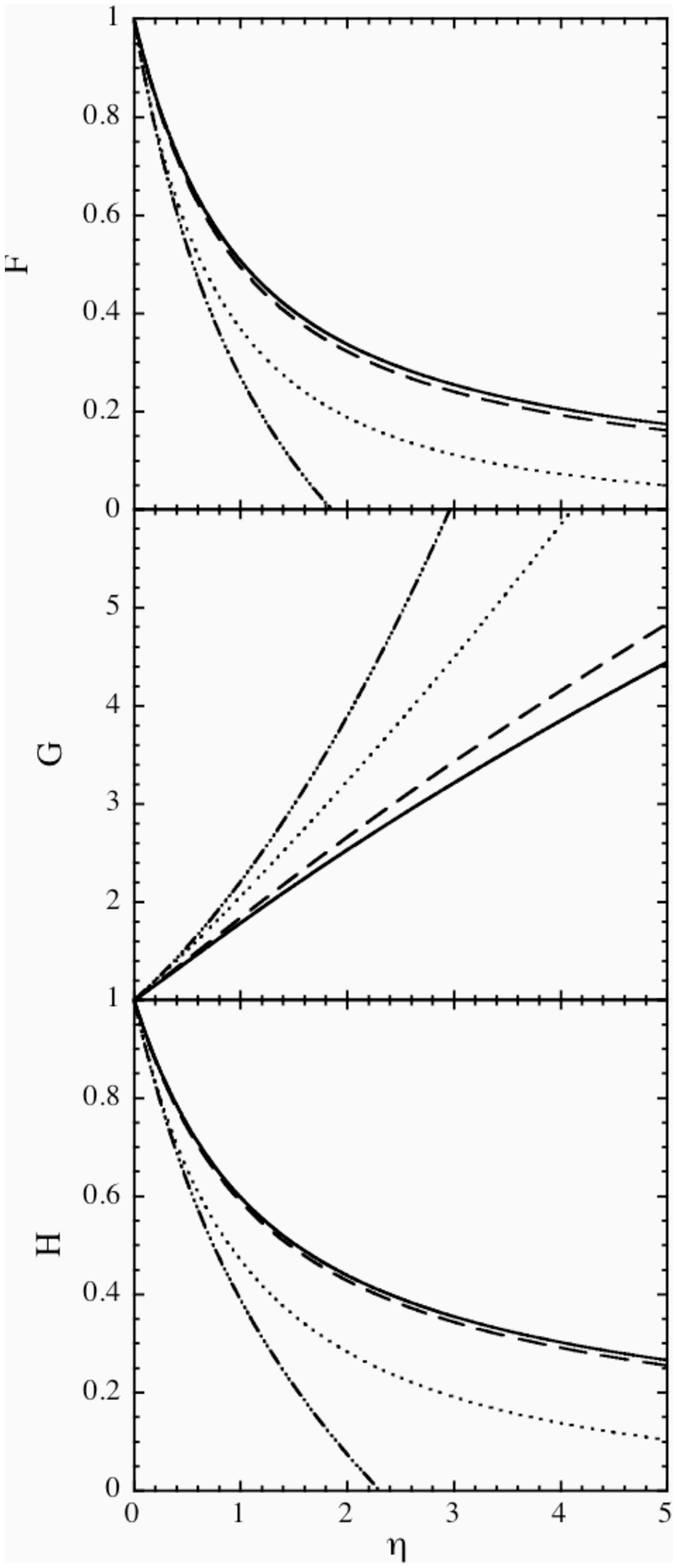}
\caption{Comparison between a planar flow and a spherical flow. The dashed, the dash-dotted, the solid and the dotted lines correspond to the self-similar solutions, the solutions involving sphericity, an ultra-relativistic PP flow  from a numerical calculation and, an ultra-relativistic SS flow from a numerical calculation, respectively.  These two numerical calculations were carried out with the parameter $A=1.0\times 10^{16}$. The solid and dotted lines from the results of the numerical calculations for PP or SS are drawn by tracing one fluid element with the preshock position $x_0/R_*=4.49\times 10^{-4}$ and at shock emergence $p_1/\rho_1 = 33.1$. The sphericity correction is obtained for $\xi=1.15\times 10^{-6}$. }
\label{fig-FGH}
\end{center}
\end{figure}

\subsection{Solutions}
The results are shown in Figure \ref{fig-FGH}. 
The dash-dotted lines represent the flow involving the sphericity correction, and the dashed lines the self-similar solutions. 
The density and the pressure involving the effect of sphericity drop more steeply than in PP. 

It is appropriate to confirm the range in which the correction is valid. 
This is one of incentives to conduct numerical calculations. 
In \S \ref{NCsphericity}, the validity of our correction will be examined. 

\section{MILDLY RELATIVISTIC FLOW}\label{MR}
\subsection{Formulation}
In later phases, $p$ becomes comparable to $\rho$. 
We will not be able to ignore the rest mass energy density in the EOS. 
We should use the EOS $\varepsilon=3p+\rho$ instead of $\varepsilon=3p$. 
In order to include the term of the rest mass energy density, we  replace the dependent variable $\rho$ in equations (\ref{beq1})-(\ref{beq3}) with the enthalpy per unit volume : $w=4p+\rho$. 
As a result, the basic equations are converted to 
\begin{equation}
	\frac{\partial}{\partial t}(w\gamma^2\beta)+
	\frac{\partial}{\partial x}(w\gamma^2\beta^2+p) =0 ,
\end{equation}
\begin{equation}
	\frac{\partial}{\partial t}(w\gamma^2-p)+
	\frac{\partial}{\partial x}(w\gamma^2\beta) =0 ,
\end{equation}
\begin{equation}
	\frac{\partial}{\partial t}[\gamma(w-4p)]+\frac{\partial}{\partial x}[\beta\gamma(w-4p)]=0 .
\end{equation}
These equations are combined to yield
\begin{equation}
	\frac{w}{2}\frac{d\gamma^2}{dt}+\gamma^2\frac{dp}{dt}=
	\frac{\partial p}{\partial t},\label{eqnss1}
\end{equation}
\begin{equation}
	\frac{1}{2}\frac{d\ln\gamma^2}{dt}+\frac{d\ln w}{dt}-\frac{1}{w}\frac{dp}{dt}
	=-\frac{\partial\beta}{\partial x},
\end{equation}
\begin{equation}
	\frac{d}{dt}[\ln(w-4p)]+\frac{1}{w}\frac{d}{dt}(p-w)=0.
\end{equation}
To use the Lagrangian coordinate as in the self-similar ultra-relativistic PP solution, another equation not containing $\rho$ is needed. This equation can be obtained by differentiating the definition of the fluid velocity with respect $a$ as
\begin{equation}
	\frac{d}{dt}\frac{\partial x}{\partial a}=\frac{\partial v}{\partial a}.
	\label{eqnss4}
\end{equation}
We seek a solution in the form
\begin{equation}
	p=p_1(a)F(\eta),\label{F}
\end{equation}
\begin{equation}
	\gamma^2=\gamma^2_1(a)G(\eta),
\end{equation}
\begin{equation}
	w=4p_1(a)W(\eta),
\end{equation}
\begin{equation}
	\frac{\partial x}{\partial a} = J(\eta),\label{J}
\end{equation}
by analogy with the ultra-relativistic case. 
We substitute (\ref{F})-(\ref{J}) into (\ref{eqnss1})-(\ref{eqnss4}) and obtain non-dimensional equations,
\begin{equation}
	\frac{2W}{F}\frac{G'}{G}+\left[1+\frac{\eta}{(m+1)GJ}\right]\frac{F'}{F}
	=\frac{\delta-m}{m+1}\frac{1}{GJ}, \label{eqmr1}
\end{equation}
\begin{equation}
	2\left(1+\frac{\eta}{m+1}\frac{1}{GJ}\right)\frac{G'}{G}
	+4\frac{W'}{W}-\frac{F'}{W}=-\frac{2m}{m+1}\frac{1}{GJ},
\end{equation}
\begin{equation}
	4\frac{W'}{W}=\left(3+\frac{F}{W}\right)\frac{F'}{F},\label{eqmr3}
\end{equation}
\begin{equation}
	J'=\frac{1}{2(m+1)G}\left(\eta\frac{G'}{G}+m\right).\label{eqmr4}
\end{equation}
The initial conditions are
\begin{eqnarray}
	F(0)  &=& 1,	\\
	G(0)  &=&1, \\
	W(0) &=& 1+\frac{\rho_1(a)}{4p_1(a)}	,\label{ICW}\\
	J(0)   &=& 1.
\end{eqnarray}	
If $W(0)=1$, the solutions agree with \cite{NS05}. 
Otherwise, they are not self-similar and $F, G, W,$ and $J$ depend not only on $\eta$ but also on $a$ and $A$ through the initial condition (\ref{ICW}). 
We will check the validity of the assumed form of the solution by comparison with results obtained from direct numerical integrations of relativistic hydrodynamical equations in \S \ref{NC}. 

\subsection{Solutions}
\begin{figure}
\begin{center}
\includegraphics[clip=true, scale=0.6]{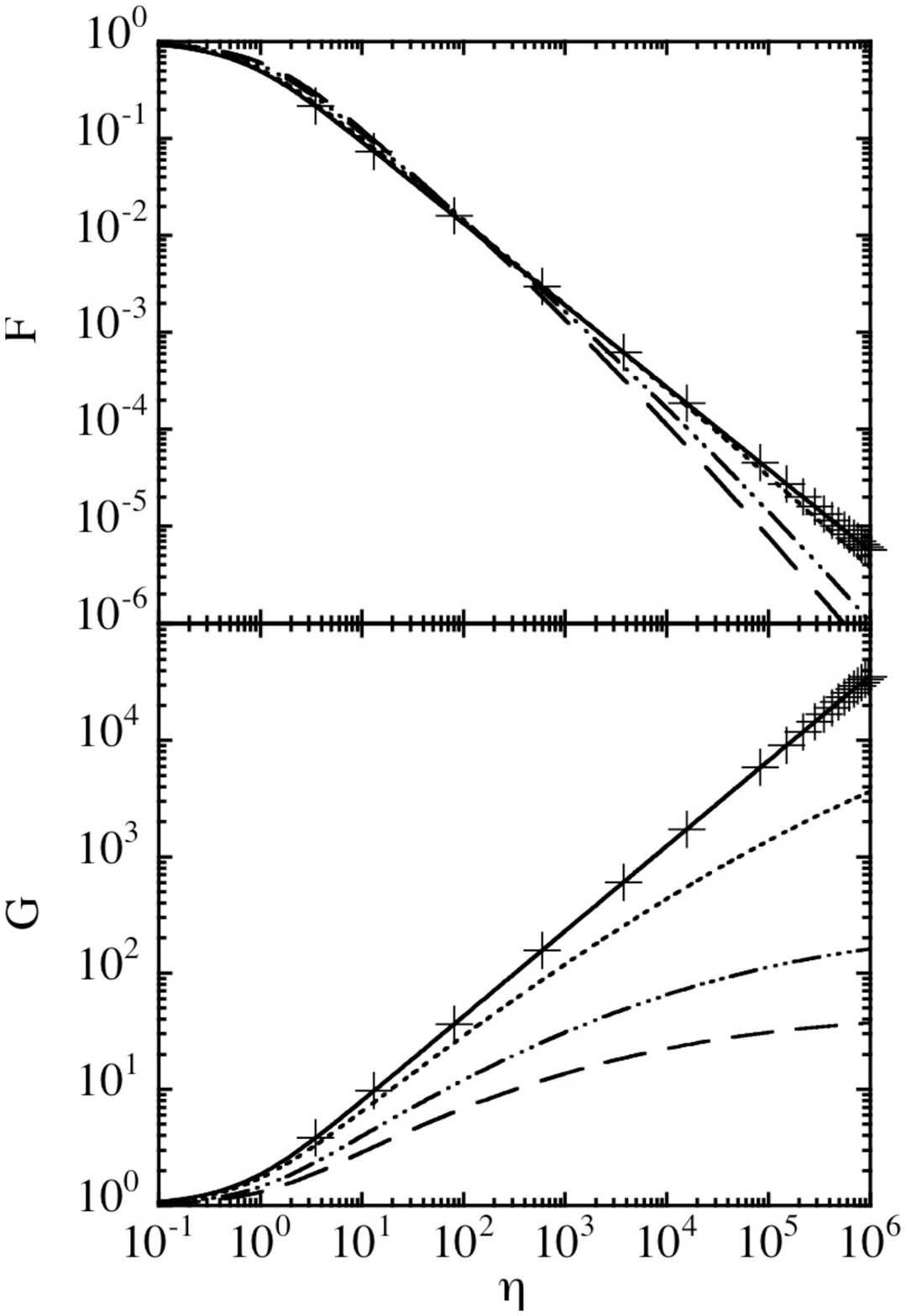}
\caption{Comparison between the semi-analytic solutions for eqs. (\ref{eqmr1})-(\ref{eqmr4}) and the self-similar solutions. The solid, the dotted, the  dash-dotted, and the dashed lines represent the semi-analytic solutions with the initial conditions $p_1/\rho_1=\infty, 2.5, 0.5, 0.25$ respectively. The crosses, $+$, represent the self-similar solutions.}
\label{fig-mrFGJ}
\end{center}
\end{figure}

$F$ and $G$ integrated with various $W(0)$ are shown in Figure \ref{fig-mrFGJ}. 
It is obvious that mildly relativistic flows, unlike an ultra-relativistic flow that forgets its history, depend on the initial energy distributions. 

From the numerical integration, it is found that $F, G, W$ and $J$ undergo power-law evolution in the limit of $\eta\to\infty$. 
The indices are shown in Table \ref{mr-indices} for $\eta > 10^6$.  
The values of $d\ln G/d\ln\eta$ in this table indicate that even though the semi-analytic solutions include the rest mass energy in the EOS, the acceleration does not terminate completely. 
For $p_1/\rho_1\ltsim 1$, the acceleration almost terminates for $\eta\to\infty$. In contrast, the acceleration does not terminate for $p_1/\rho_1\gtsim 1$ . 

\begin{deluxetable}{lcccc}
\tabletypesize{\scriptsize}
\tablecaption{Power-law indices of $F, G, W$ and $J$ for the semi-analytic  solutions\label{mr-indices}}
\tablewidth{0pt}
\tablehead{
 \colhead{$p_1/\rho_1$} &
 \colhead{$d\ln F/d\ln\eta$} & 
 \colhead{$d\ln G/d\ln\eta$} & 
\colhead{$d\ln W/d\ln\eta$}  &
\colhead{$d\ln J/d\ln\eta$}  
}
\startdata
$\infty$   & $-0.845$
	& 0.732
	& $-0.845$
	& 0.268		\\
2.5   & $-1.01$
	& 0.304
	& $-0.813$
	& 0.607 \\
1.0   & $-1.13$
	& 0.159
	& $-0.8744$	
	& 0.772\\	
0.5  & $-1.24$
	& 0.07
	& $-0.935$
	& 0.893 \\
0.25 & $-1.27$
	& 0.04
	& $-0.957$
	& 0.931 	
 \enddata

\end{deluxetable}

\section{NUMERICAL CALCULATIONS}
\label{NC}
\subsection{Calculation procedure}
We carry out numerical calculations with the Godunov method for relativistic hydrodynamics in the Lagrangian coordinate system \citep{Marti95}. 
To write down the basic equations for relativistic hydrodynamics in conservation form, we have to introduce some new variables $V, s,$ and $Q$.
These variables are related to quantities $\rho$ and $w$ in the local rest frame of the fluid through
\begin{equation}
	V \equiv 1/(\rho\gamma), 
\end{equation}
\begin{equation}
	s \equiv (w\gamma v)/\rho, 
\end{equation}
\begin{equation}
	Q = (w\gamma^2-p)/(\rho\gamma)-1.
\end{equation}
For PP geometry, the basic equations are expressed with these variables as
\begin{eqnarray}
	\frac{dV}{dt} &=& \frac{\partial v}{\partial m_x}, \\
	\frac{ds}{dt}  &=& - \frac{\partial p}{\partial m_x}, \\
	\frac{dQ}{dt} &=& - \frac{\partial (pv)}{\partial m_x},
\end{eqnarray}
where the Lagrangian coordinate is defined as
\begin{equation}
	dm_x=-\rho_1\gamma_1da=-\rho\gamma dx.
\end{equation}
For SS geometry, 
\begin{eqnarray}
	\frac{dV}{dt} &=& \frac{\partial}{\partial m_r}(4\pi r^2v),\\
	\frac{ds}{dt}  &=& - \frac{\partial}{\partial m_r}(4\pi r^2p), \\
	\frac{dQ}{dt} &=& - \frac{\partial}{\partial m_r}(4\pi r^2pv),
\end{eqnarray}
where the Lagrangian coordinate is defined as
\begin{equation}
	dm_r = -4\pi (R_*-a)^2\rho_1\gamma_1da = 4\pi r^2\rho\gamma dr.
\end{equation}
These equations are integrated numerically by evaluating the derivatives in the right hand side using the Godunov method. 
The difference lies in the EOS; $\varepsilon = 3p+\rho$ in numerical calculations while $\varepsilon = 3p$ in the self-similar solutions.   

Calculations were carried out for both PP and SS geometry,  
from the moment of shock emergence ($t=0$). 
We set the initial condition from the self-similar solutions at the time $t=0$. 
The parameters used for calculations are $\delta=3$, $A=1.0\times10^{16}$, $1.0\times 10^{12}$ or $1.0\times 10^{10}$, and $R_*=1.0\times10^9$ m. Here $A$ has a dimension of some power of time. Thus the values of $A$ are expressed using second as a unit of time.
Note that $\gamma_1$ depends on $A$ as
\begin{equation}
	\gamma_1\propto A^{1/2(m+1)}\sim A^{0.2}. 
\end{equation}
Hence $\gamma_1$ varies by a factor of about 10 for the above values of $A$. 

We obtain $p$, $\rho$ and $\gamma$ directly from numerical results. 
The non-dimensional variables $F, G,$ and $H$ can be obtained by $p/p_1$, $\gamma/\gamma_1$, and $\rho/\rho_1$, respectively.  
On the other hand, when we made calculations starting before shock emergence, the moment of shock emergence could not be determined exactly. As a result we could not determine $p_1$,  $\gamma_1$, or $\rho_1$ to sufficient accuracy.
This was caused by the finite thickness of the shock front in numerical calculations. 
Therefore, we dedicate ourselves to calculations after shock emergence.

\subsection{Results}
\label{NCresults}
\subsubsection{Connection to the initial states}
In presenting the results after shock emergence, it is preferable to describe those in terms of the original position of each fluid element $x_0$ rather than the Lagrangian coordinate $a$ to help actual applications to astrophysical phenomena. 
For PP, the definition of the Lagrangian coordinate provides the exact relation between $a$ and $x_0$.   
\begin{equation}
	\int_0^{x_0}\rho_0 dx_0 = \int_0^a\rho'_1da,
\end{equation}
Substituting $\rho_0=bx_0^\delta$ and the expression for $\rho'_1$ (Eq. (\ref{density@sb})) into both sides of this equation, we obtain
\begin{eqnarray}
	x_0 &=&\left[2A(m+1)\right]^{1/(\delta+1)} \nonumber\\ 
	&&\times\left(\frac{2A[1+(2\sqrt{3}-3)\delta]}{[3+2\sqrt{3}(1+\delta)]
	^{\sqrt{3}(1+\delta)/\{2[1+\sqrt{3}(1+\delta)]\}}}
	\right)^{\frac{(\delta-m)}{(m+1)(\delta+1)}}\nonumber \\
	&&\times a^{1/(m+1)}.
\end{eqnarray}
If $\delta=3$, then $x_0=1.56\times A^{0.418}\times a^{0.418}$. Though the corresponding relation $r_0=r_0(a)$ for SS will be obtained from
\begin{equation}
	\int_0^a4\pi(R_*-a)^2\rho'_1da=\int_{r_0}^{R_*}4\pi r_0^2\rho_0 dr_0,
\end{equation}
 it is difficult to express $r_0$ as a function of $a$ in an analytic form unless we truncate higher terms of $a/R_*$ and $x_0/R_*$. To the lowest order, it takes the same form  as PP,
 \begin{eqnarray}
	r_0&\approx & \left[2A(m+1)\right]^{1/(\delta+1)}\nonumber\\
	&&\times\left(\frac{2A[1+(2\sqrt{3}-3)\delta]}{[3+2\sqrt{3}(1+\delta)]
	^{\sqrt{3}(1+\delta)/\{2[1+\sqrt{3}(1+\delta)]\}}}
	\right)^{\frac{(\delta-m)}{(m+1)(\delta+1)}}\nonumber \\
	&&\times a^{1/(m+1)},
\end{eqnarray}
Using these relations, a fluid element after shock emergence can be traced back to its original position.

In addition, we can see how the parameters characterizing the explosion such as the injected energy  $E_\mathrm{inj}$ and the ejected mass $M_\mathrm{ejc}$ affect the flow through the parameter $A$ as follows. The parameter $A$ can be related to $E_\mathrm{inj}$ and $M_\mathrm{ejc}$ by 
\begin{eqnarray}
	E_\mathrm{inj}&=&\int \gamma^2\rho dx , \nonumber\\
		   &\propto& \Gamma^2 (-t)^{-m}M_\mathrm{ejc} , \nonumber\\
		   &\propto	& AM_\mathrm{ejc} .
\end{eqnarray}
Thus we obtain 
\begin{equation}
	A\propto \frac{E_\mathrm{inj}}{M_\mathrm{ejc}}. \label{A}
\end{equation}
It is desirable to get the proportional constant of (\ref{A}) which connects the shock velocity with ${E_\mathrm{inj}}$ and ${M_\mathrm{ejc}}$. 
\cite{Tan00} proposed an analytical formula (eqns (4) and (9) in their paper) to estimate such a coefficient.  See also equations (17), (18) and (19) in \cite{Matzner99}. 
In our case, we can obtain an expression of $A$ as a function of  ${E_\mathrm{inj}}$ and ${M_\mathrm{ejc}}$ by substituting $\rho=b(-t)^\delta$ into equation (4) in \citet{Tan00}. 

\subsubsection{Effects of sphericity}\label{NCsphericity} 
We discuss the validity of our sphericity correction in this section. 
Figure \ref{fig-FGH} shows the trajectories of the fluid elements with the preshock position $x_0/R_*=4.49\times 10^{-4}$ and at shock emergence  $p_1/\rho_1=33.1$ (The ratio of the internal energy to the kinetic energy is about 132). 
The solutions of (\ref{solF})-(\ref{solH}) including sphericity correction (the dash-dotted line) match the numerical results for SS with $A=1.0\times 10^{16}$ (the dotted line) for small $\eta$.
On the contrary, the planar flows start to deviate from the beginning.  
The degree of accuracy in the sphericity correction is shown in Table \ref{table-error}. 
Our description for an SS flow is a good approximation in the range of $\eta \ltsim 0.5$ within accuracy of 5 \% . 

\begin{deluxetable}{lccc}
\tabletypesize{\scriptsize}
\tablecaption{Differences between the numerical solutions for a spherical flow and the self-similar solutions with correction terms for sphericity\label{table-error}}
\tablewidth{0pt}
\tablehead{
 \colhead{$\eta$} &
 \colhead{$\Delta F/F \ (\%)$} & 
 \colhead{$\Delta G/G \ (\%)$} & 
\colhead{$\Delta H/H \ (\%)$}  
}
\startdata
0.3   & 2.1
	& 1.0
	& 1.2		\\
0.4   & 3.9
	& 1.8
	& 2.4 \\
0.5  	& 6.1
	& 2.5
	& 3.9 
 \enddata

\end{deluxetable}

\begin{figure}
\begin{center}
\includegraphics[clip=true,scale=0.55]{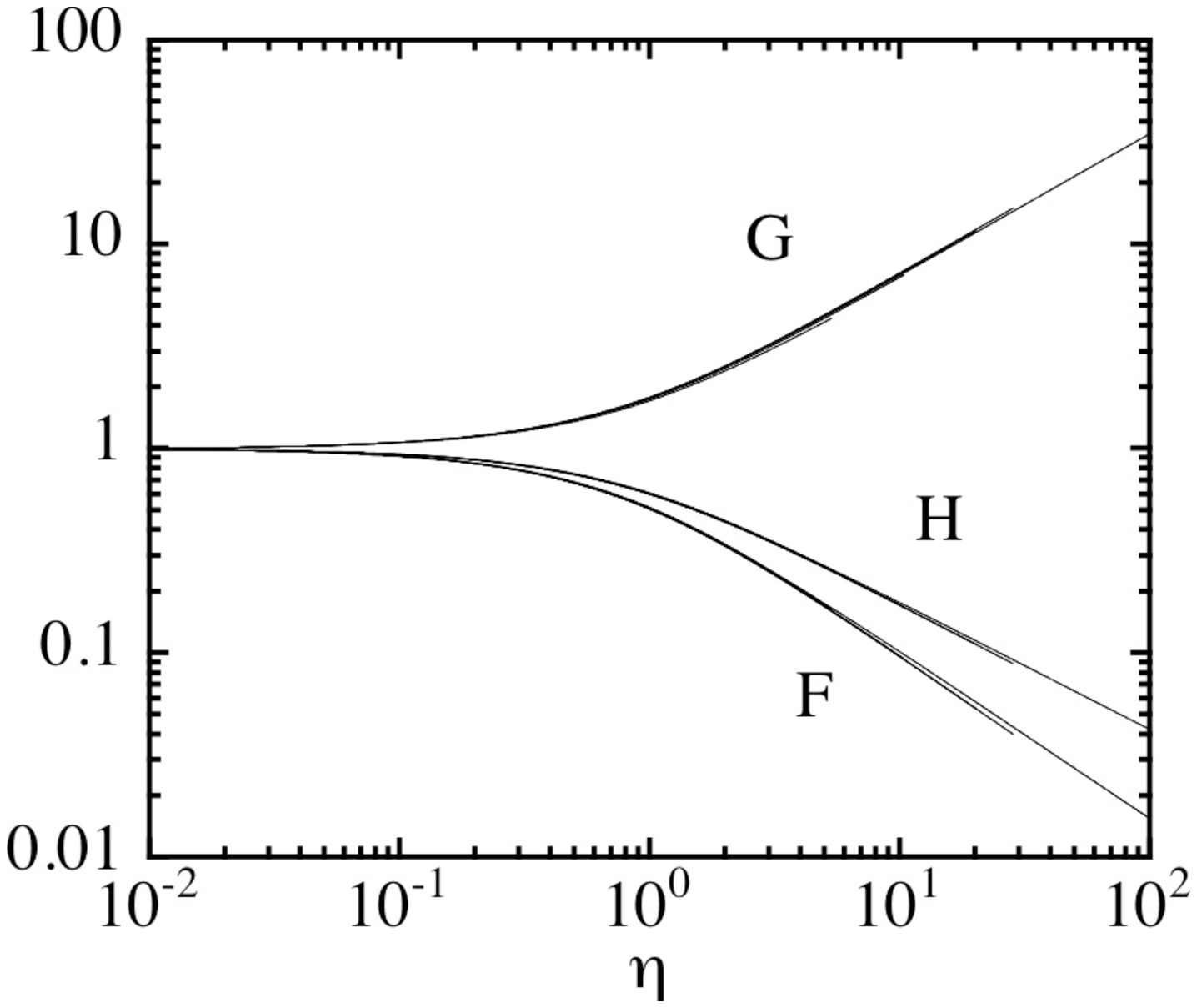}
\caption{Numerical results of an ultra-relativistic PP flow for $A=1.0\times 10^{16}$.  Time evolutions of F, G and H are shown. Each line represents the locus of one fluid elements with the preshock position from $x_0/R_*=6.27\times 10^{-4}$ to $1.36\times 10^{-2}$ and at shock emergence $p_1/\rho_1$ ranging from $3.09$ to $26.3$. A considerable overlap of all the lines suggests self-similarity of the flow.}
\label{fig-NPFGH}
\end{center}
\end{figure}

Next, we compare an ultra-relativistic PP and SS flow from the results of numerical calculations shown in Figure \ref{fig-FGH}. 
The density of SS decreases more drastically than that in  PP (the solid lines).  
This results from the difference in the geometry; the volume of a spherical shell expands more than that in a PP shell. 
Then, the pressure in SS also decreases more than that in PP, because the flow is adiabatic. 
The Lorentz factor in SS increases more enormously than that in PP. 
As the acceleration is advanced, the ratio of the internal energy to the kinetic energy which is about $4p/\rho$ for $\gamma\to\infty$ in the observing frame drops with decreasing $\rho$ because $p/\rho$ decreases as $\rho^{1/3}$ in an adiabatic flow.   
As a consequence, the acceleration of an SS flow is more quickly advanced and terminates in a shorter time scale than PP. 

\subsubsection{Intermediate distribution}

\begin{figure}
\begin{center}
\includegraphics[clip=true,scale=0.85]{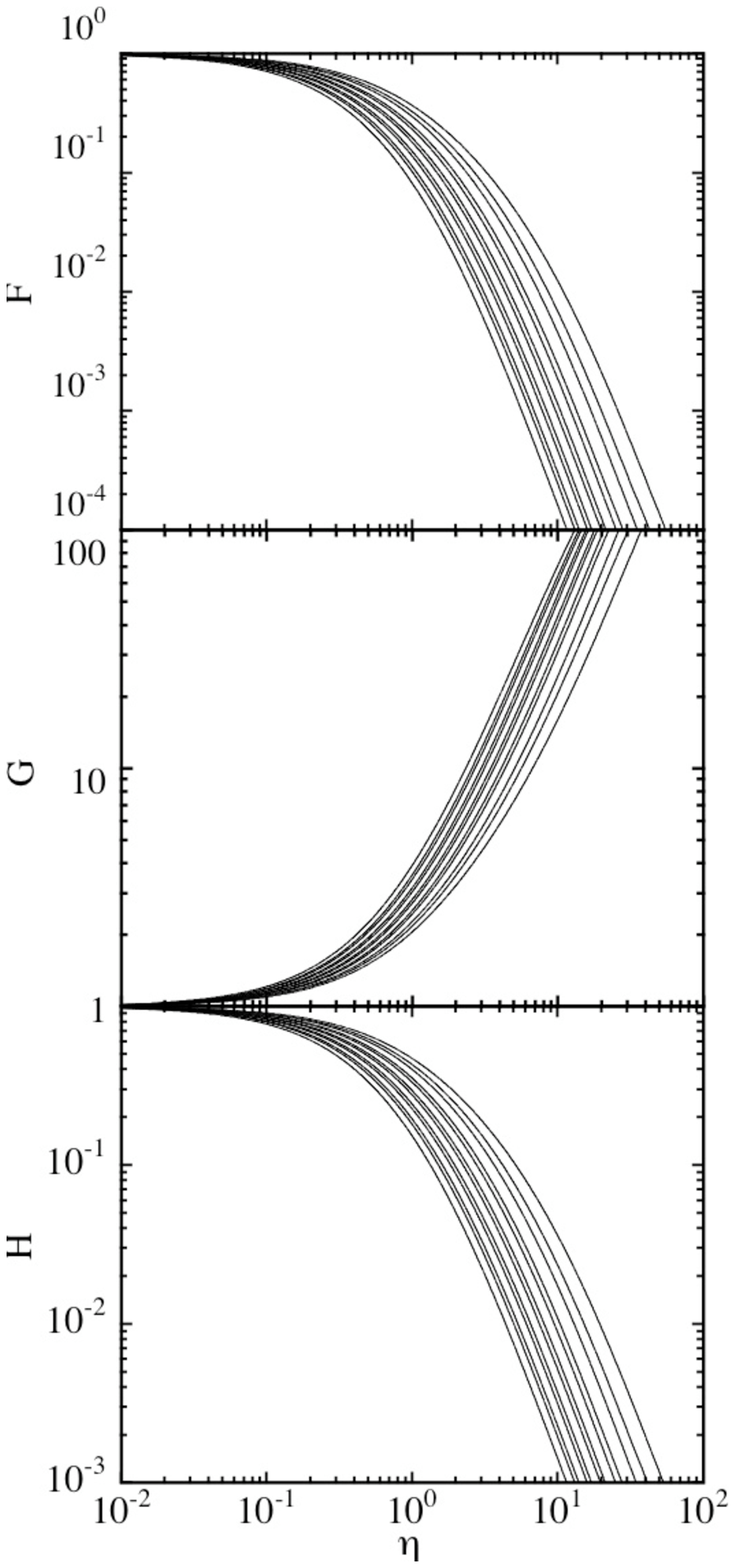}
\caption{Same as Figure \ref{fig-NPFGH} but for a UR spherical flow with the preshock position from $x_0/R_*=4.49\times 10^{-4}$ to $3.29\times 10^{-3}$ and at shock emergence $p_1/\rho_1$ ranging from $8.48$ to $33.1$. }
\label{fig-NSFGH}
\end{center}
\end{figure}

For $\eta\to\infty$, $F$, $G$, and $H$ from the self-similar solution have power law distributions with respect to $\eta$. 
We call such distributions as the intermediate distributions.  
$F, G, \ \mathrm{and} \ H$ from numerical calculations can be fitted  by power law distributions. 
The indices are shown in Table \ref{table-indices} which are obtained for $\eta\sim 10-100$. 
Comparing the self-similar flow and the ultra-relativistic PP flow obtained from numerical calculations, the differences of indices between them are 1.1\%, 5.7\%, and 0.63\% for $d\ln F/d\ln\eta$, $d\ln G/d\ln\eta$, and $d\ln H/d\ln\eta$, respectively. 
The difference in $d\ln G/d\ln\eta$ is particularly prominent, which implies that the effect of the rest mass energy density shows up to suppress the acceleration. 
That is to say, $d\ln G/d\ln\eta$ of numerical calculations is smaller than that of the self-similar solutions. The values of $p_1/\rho_1$ ranges form 10 to 70 depending on $a$ in this particular numerical calculation. 

Figure \ref{fig-NPFGH} shows the evolution of several fluid elements in a PP flow obtained from a numerical calculation. 
Though each line has a different value of $p_1/\rho_1$ in the range from $8.5$ to $33$, the trajectories are insensitive to it, which indicates the flow is very close to self-similar.
If we set $p_1/\rho_1\to \infty$, the result will approach the self-similar solutions.  

Numerical results of an SS flow  behave differently from that of a PP flow. 
These are shown in Figure \ref{fig-NSFGH}.
The trajectory of each fluid element of an SS flow depends on the initial position $a$. 
It is insufficient to use $\eta$ for characterizing the flow. 
The values of indices change in the ranges shown in Table \ref{table-indices}

\begin{deluxetable}{lccc}
\tabletypesize{\scriptsize}
\tablecaption{Power-law indices of $F, G,$ and $H$  for ultra-relativistic flows in the intermediate asymptotic phase \label{table-indices}}
\tablewidth{0pt}
\tablehead{
 \colhead{} &
 \colhead{$d\ln F/d\ln\eta$} & 
 \colhead{$d\ln G/d\ln\eta$} & 
\colhead{$d\ln H/d\ln\eta$}  
}
\startdata
self-similar   & $-0.845$
	& 0.732
	& $-0.634$		\\
PP (numerical)   & $-0.836$
	& 0.690
	& $-0.630$ \\
SS (numerical)  	& $-3.23$ - $-2.97$
	& 1.33 - 1.44
	&$-2.43$ - $-2.27$
 \enddata

\end{deluxetable}

\subsubsection{Mildly relativistic flow}
\begin{figure}
\begin{center}
\includegraphics[clip=true,scale=0.55]{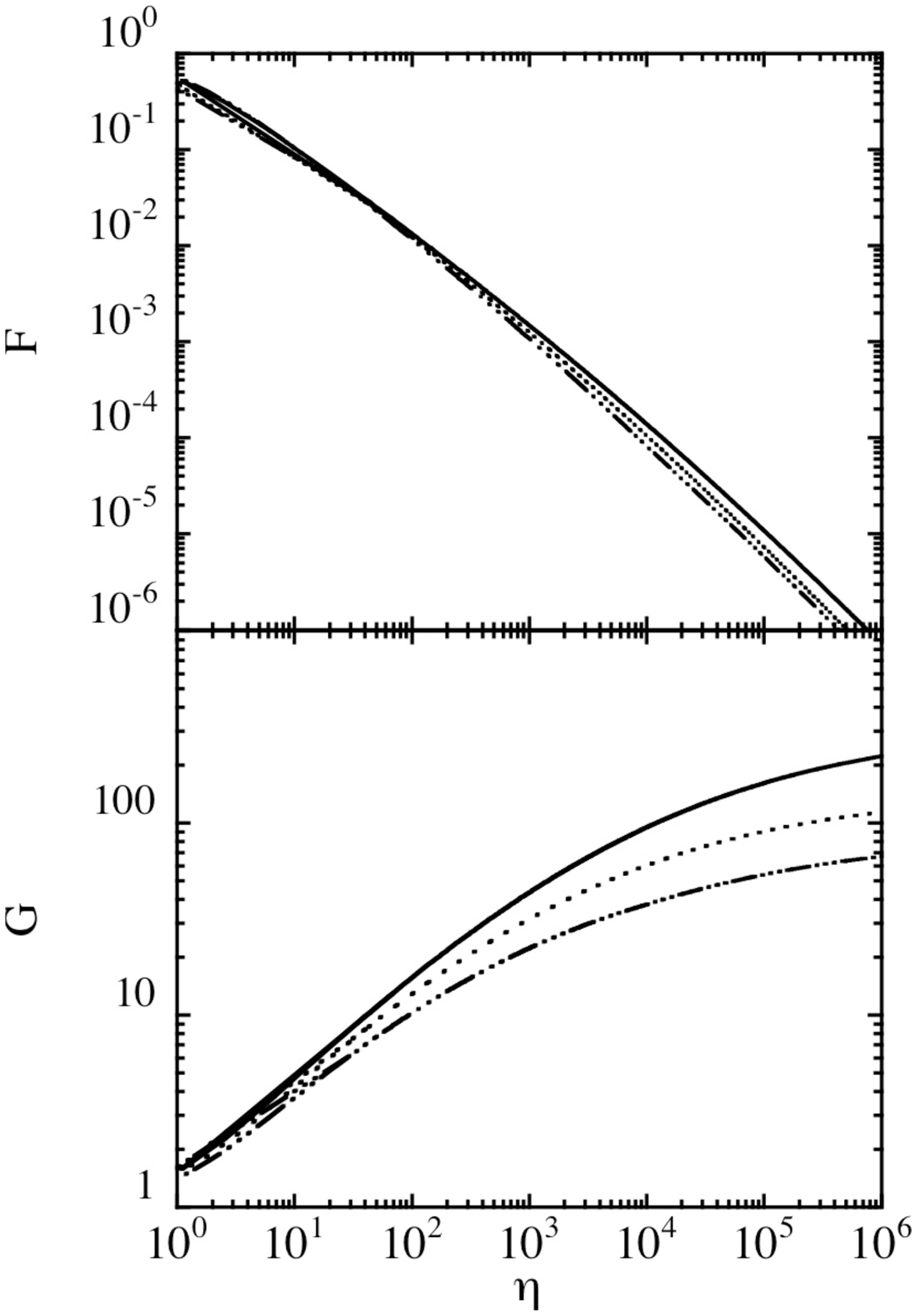}
\caption{Numerical results for a mildly relativistic PP flow. This calculation was carried out with $A=1\times 10^{10}$. 
The solid, the dotted, and the dash-dotted lines correspond to the fluid element with the preshock initial position, $x_0/R_*= 1.05\times 10^{-3}, \ 1.72\times 10^{-3}, \ 2.88\times 10^{-3}$, and at shock emergence the strength of relativity,  $p_1/\rho_1=1.022,\, 0.727,\, 0.507$, respectively.}
\label{fig-lowEpFGH}
\end{center}
\end{figure}

\begin{figure}
\begin{center}
\includegraphics[clip=true,scale=0.55]{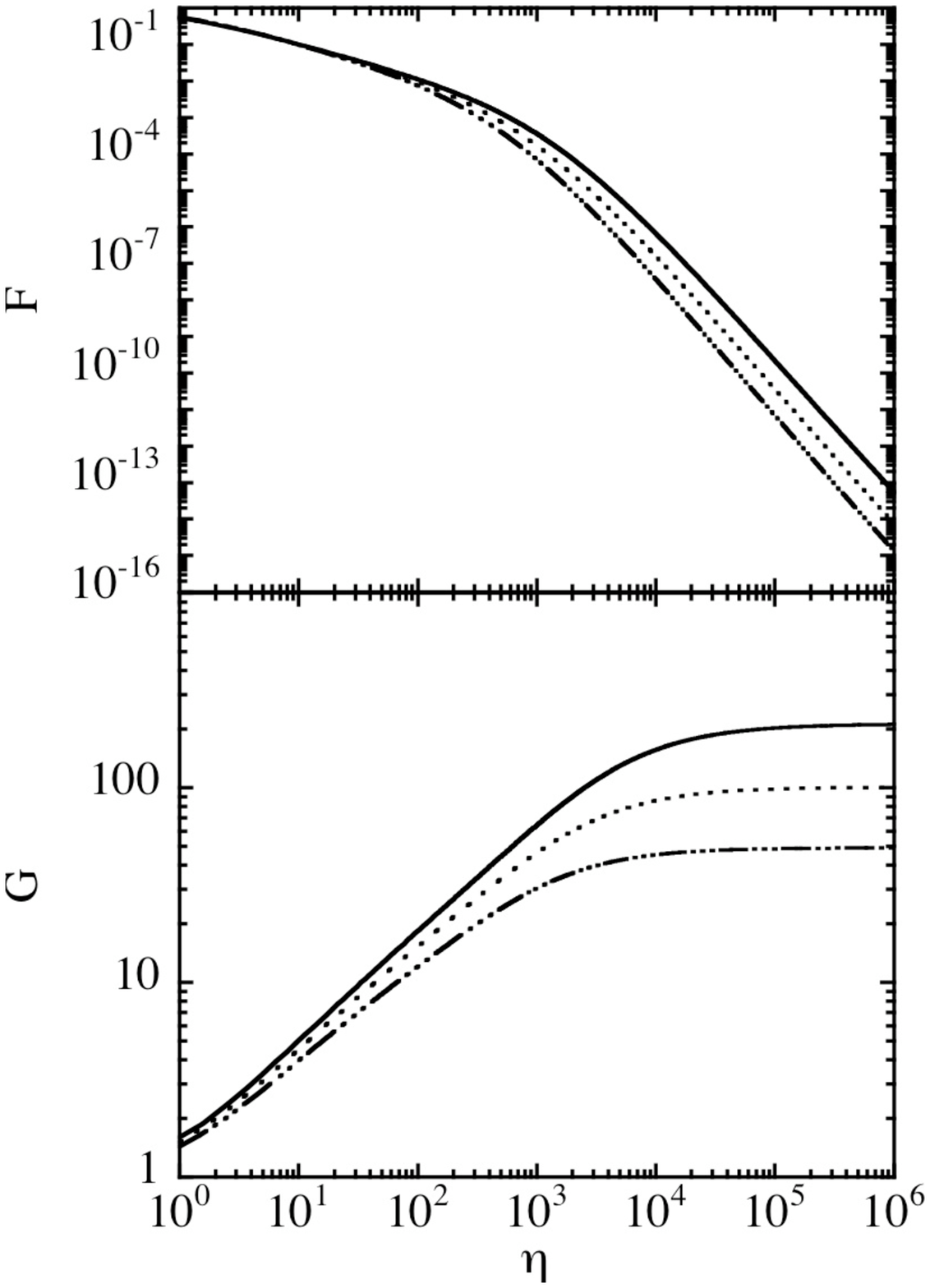}
\caption{Numerical results for a mildly relativistic SS flow. This calculation was carried out with $A=1\times 10^{10}$. 
The solid, the dotted, and the dash-dotted lines correspond to a fluid element with the preshock initial position, $x_0/R_*= 7.96\times 10^{-4}, \ 1.33\times 10^{-3}, \ 2.37\times 10^{-3}$, and at shock emergence the strength of relativity,  $p_1/\rho_1=1.24,\, 0.870,\, 0.580$, respectively.}
\label{fig-lowEsFGH}
\end{center}
\end{figure}

The acceleration in numerical calculations is suppressed compared to that of the self-similar solutions due to the rest mass energy in the EOS used in numerical calculations. 

\begin{figure}
\begin{center}
\includegraphics[clip=true,scale=0.55]{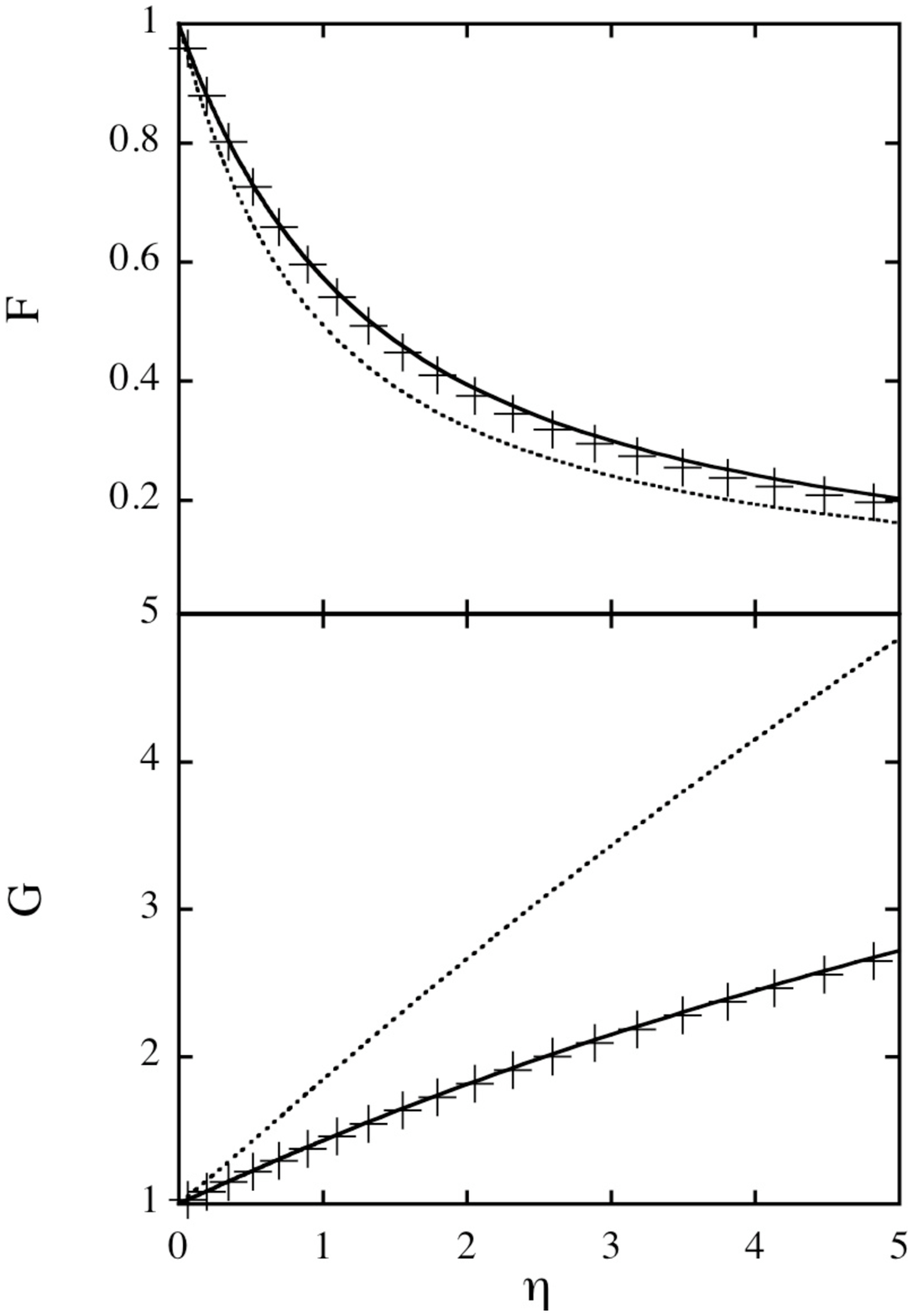}
\caption{Comparison between the self-similar solutions, the semi-analytic solutions for a mildly relativistic PP flow, and the numerical results for a mildly relativistic PP flow.  The dotted lines, the solid lines and the crosses, $+$, represent the self-similar solutions, the semi-analytic solutions, and the numerical results,  respectively. This numerical calculation was carried out with $A=1.0\times 10^{10}$. 
And the crosses represents a trajectory of a fluid element with $x_0/R_*=0.00325$ and $p_1/\rho_1=0.466$. }
\label{fig-compMRFG1}
\end{center}
\end{figure}

\begin{figure}
\begin{center}
\includegraphics[clip=true,scale=0.55]{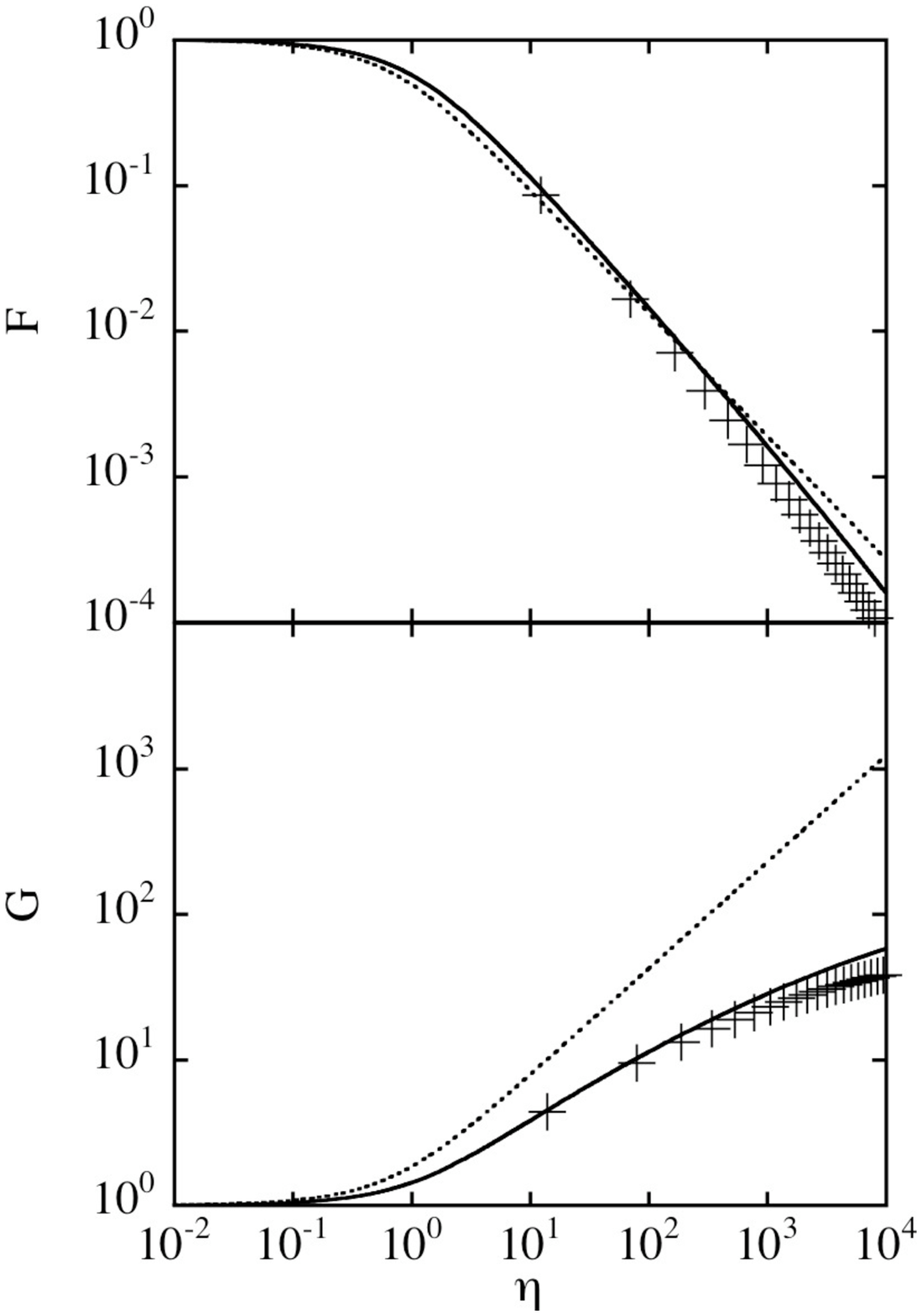}
\caption{Same as Figure \ref{fig-compMRFG1} but for a more extended range of $\eta$.}
\label{fig-compMRFG2}
\end{center}
\end{figure}

The results of mildly relativistic flows from numerical calculations are shown in Figure \ref{fig-lowEpFGH} for a PP configuration and Figure \ref{fig-lowEsFGH} for an SS configuration. 
In Figures \ref{fig-compMRFG1} and \ref{fig-compMRFG2}, we compare the semi-analytic solutions (PP) for a mildly relativistic flow with the corresponding results from numerical calculations.  
The power law indices in the intermediate distribution are shown in Table \ref{table-mr-indices}. 
The semi-analytic solutions give a precise description of a mildly relativistic flow up to $\eta\sim 100$. 

\begin{deluxetable}{lccccc}
\tabletypesize{\scriptsize}
\tablecaption{Power-law indices of $F$, $G$, and $H$  for mildly  relativistic flows in the final free expansion phase (from numerical results)\label{table-mr-indices}}
\tablewidth{0pt}
\tablehead{
 \colhead{} &
 \colhead{$x_0/R_*$} &
 \colhead{$p_1/\rho_1$} &
 \colhead{$d\ln F/d\ln\eta$} & 
 \colhead{$d\ln G/d\ln\eta$} &
 \colhead{$d\ln H/d\ln\eta$}
}
\startdata
PP  
	& $1.05\times 10^{-3}$
	& 1.02
	& $-1.21$
	& 0.03		
   	& $-$0.98\\
	& $1.72\times 10^{-3}$   
	& 0.727
	& $-1.25$
	& 0.02
	& $-$0.98			\\
	& $2.88\times 10^{-3}$
	& 0.507
	& $-1.26$
	& 0.03
	   & $-$0.97 \\
SS 
	& $7.96\times 10^{-4}$
	& 1.24
	& $-3.31$
	& 0.00		
	& $-2.99$\\
	& $1.54\times 10^{-3}$
	& 0.782
	& $-3.45$
	& 0.00			
        &$-2.99$\\
        & $2.59\times 10^{-3}$
	& 0.546
	& $-3.52$
	& 0.00 
	   &$-3.00$
\enddata

\end{deluxetable}

\subsubsection{Final distribution}
We investigate the final energy distributions for the mildly relativistic flow and the PP flow from numerical calculations.
The energy per unit mass excluding the rest-mass energy is given by 
\begin{equation}
	\epsilon=\frac{\tau}{\rho\gamma},
\end{equation}
where
\begin{equation}
	\tau=(\rho+4p)\gamma^2-p-\rho\gamma.
\end{equation}
In the stage of free expansion, the ejecta are accelerated to $\gamma=\gamma_\mathrm{f} \gg 1$. Thus, $\gamma_\mathrm{f}\sim\epsilon$.
Using the definition of the Lagrangian coordinate, we calculate the energy spectrum defined as the total mass $M(>\gamma_\mathrm{f})$ with the energy per unit mass greater than $\gamma_\mathrm{f}\sim\epsilon$ as
\begin{eqnarray}
	M(>\gamma_\mathrm{f}) &=&\int\rho\gamma_\mathrm{f} dx , \nonumber \\
				&\propto& b A^{(\delta+1)/(m+1)}\gamma_\mathrm{f}^{-2.6} .
\end{eqnarray}
Furthermore, it is useful to present the energy of the ejecta $E(>\gamma)$ with  Lorentz factors greater than $\gamma$. From the numerical results for a PP mildly relativistic flow, when $\eta\to\infty$ we obtain the relation between $a$ and the final Lorentz factor $\gamma_\mathrm{f}$, 
\begin{equation}
	\gamma_\mathrm{f} \propto a^{-0.64}.
\end{equation}
Using this relation, we obtain 
\begin{eqnarray}
	E(>\gamma) &=&\int\gamma_\mathrm{f} \rho\gamma_\mathrm{f} dx ,\nonumber\\
		             &=&\int\rho_1\gamma_1\gamma_\mathrm{f} da ,\nonumber\\
		             &\propto& bA^\frac{2+(m+1)\delta}{2(m+1)}\gamma_\mathrm{f}^{-1.6} .
\end{eqnarray}

\citet{Tan00} derived an empirical formula for the final Lorentz factor $\gamma_\mathrm{f}$ of each element as a function of  important parameters such as $\rho_0, E_\mathrm{inj}, M_\mathrm{ejc}, x$ that characterize the shock emergence phenomenon.  For a PP flow, They obtained
\begin{equation}
	\gamma_\mathrm{f}\sim \sqrt{1+\{[2.03+(q(1+q^2)^{0.12})^{\sqrt{3}}]q(1+q^2)^{0.12}\}^2},\label{g_fpp}
\end{equation}
where $q$ is defined as
\begin{equation}
	q=0.736\left(\frac{\tilde{E}_\mathrm{inj}}{\tilde{m}}\right)^{1/2}\left[\frac{\tilde{m}M_\mathrm{ejc}}{\rho_0(R_*-x)^3}\right]^{0.187}.
\end{equation}
Here $\tilde{m}(r)$ is the fraction of $M_\mathrm{ejc}$ within $r$ excluding any remnant mass $M_\mathrm{rem}$, i.e. $\tilde{m}(r)=[m(r)-M_{rem}]/M_\mathrm{ejc}$ . 
$\tilde{E}_\mathrm{inj}$ is the injected energy in units of the rest mass energy of $M_\mathrm{ejc}$, i.e. $\tilde{E}_{\mathrm{inj}}=E_{\mathrm{inj}}/M_{\mathrm{ejc}}$ . 
In the limit of $q\rightarrow\infty$, this equation tends to reproduce \citet{Johnson71} and thus $M(>\gamma_\mathrm{f}) \propto \gamma_\mathrm{f}^{-2.1}$, which is also derived from an ultra-relativistic self-similar solution \citep{NS05}. On the other hand, \cite{Sakurai60}'s solutions give $M(>\gamma_\mathrm{f})\propto \gamma_\mathrm{f}^{-3.6}$. These results suggest that more relativistic the shock is,  more energies can be transferred to the matter closer to the surface. 
This fact is already pointed out and formulated by \cite{Tan00}. 
See their equations, (35), (36), and (37),  \citep[see also][]{Johnson71}.
 
Next we discuss the energy spectrum for spherical flows. 
From the numerical result ($A=1\times 10^{10}$) for  the mildly relativistic SS flow, 
\begin{equation}
	\gamma_\mathrm{f} \propto a^{-0.62}. 
\end{equation}
The energy spectrum for SS is
\begin{eqnarray}
	M(>\gamma_f)&=&\int \rho\gamma_\mathrm{f} 4\pi r^2dr ,\nonumber\\
		&=&\int\rho_1\gamma_14\pi(R_*-a)^2da ,\nonumber\\
		&\sim& 4\pi R_*^2\int\rho_1\gamma_1da ,\nonumber\\
	        &\propto& b A^{(\delta+1)/(m+1)}\gamma_\mathrm{f}^{-2.7}.
\end{eqnarray}
In terms of the energy,
\begin{equation}
	E(>\gamma)\propto \gamma_\mathrm{f}^{-1.7} .
\end{equation}

\begin{figure}
\begin{center}
\includegraphics[clip=true,scale=0.5]{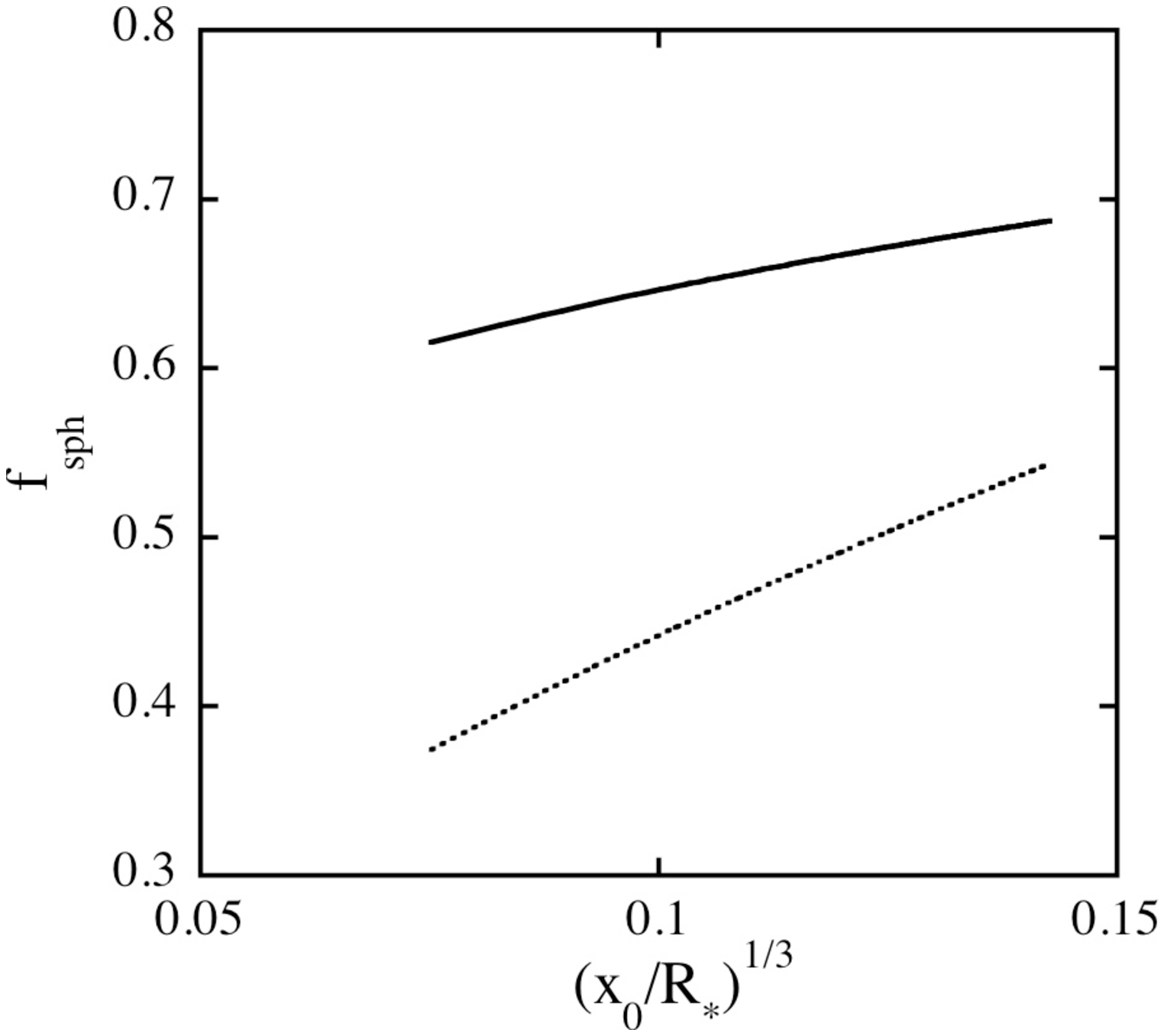}
\caption{The deficit factors $f_\mathrm{sph}$ (see text for detail). These results are  obtained by numerical calculations for both PP and SS configurations with $A=1.0\times 10^{10}$ (solid line) and $A=1.0\times10^{12}$ (dotted line). }
\label{f9}
\end{center}
\end{figure}

Although the definition of $M(>\gamma_\mathrm{f})$ in the PP configuration differs from that in the SS configuration, the difference in geometry is negligible as far as we deal with very thin shells.  
Sphericity does not significantly alter the exponents of $\gamma_\mathrm{f}$ in $E(>\gamma_\mathrm{f})$ and $M(>\gamma_\mathrm{f})$ from those of  the PP  flow, but reduces the absolute values of $E(>\gamma_\mathrm{f})$ and  $M(>\gamma_\mathrm{f})$ compared to the PP flow. 

Following \cite{Tan00}, we define $f_\mathrm{sph}$ which is a deficit factor to characterize the reduction of the final velocity of an SS flow from the corresponding PP flow,
\begin{equation}
	f_\mathrm{sph}\equiv\frac{(\gamma_\mathrm{f}\beta_\mathrm{f})_\mathrm{spherical}}{(\gamma_\mathrm{f}\beta_\mathrm{f})_\mathrm{planar}}.
\end{equation}
We show the deficit factor $f_\mathrm{sph}$ of the flows with $A=1.0\times 10^{10}$ and $A=1.0\times10^{12}$ in Figure  \ref{f9}. This figure clearly shows that the effect of sphericity is more prominent in the outer layers, which was seen in the flows with $\tilde{E}_\mathrm{inj}=0.3$ calculated by \cite{Tan00}. On the contrary, they also showed that non-relativistic SS flows become identical to the corresponding PP flows in the outer layers. This figure also indicates that the spherical expansion in a more intense explosions with larger $A$ suppress the acceleration to a greater degree. Note, however, that the PP flow with $A=1.0\times 10^{12}$ does not reach the free expansion phase in our calculation. Thus the deficit factor $f_\mathrm{sph}$ for $A=1.0\times 10^{12}$ shown in this figure will be still decreasing as time elapses.

\section{CONCLUSIONS AND DISCUSSION}
\label{CD}
We have derived the formula for sphericity correction to ultra-relativistic PP flows and presented the semi-analytic solutions for mildly relativistic flows after shock emergence. 
Both are proved to have a good agreement with numerical results at least in the earliest phases of evolution after shock emergence. 

Although the semi-analytic solutions include the rest mass energy in the EOS, this inclusion of the rest mass energy does not lead to terminating the acceleration if the pressure to rest mass energy density ratio $p_1/\rho_1$ is large enough. 
The semi-analytic solutions can not describe the whole evolution of a flow for finite values of $p_1/\rho_1$. 
This may be attributed to the assumption that a flow is described with one non-dimensional parameter except the initial condition. 
Reaching the stage of free expansion, a PP flow appears to deviate from self-similarity even if it is ultra-relativistic. 
Furthermore we neglect the effect of radiative diffusion which plays the critical role to suppress the acceleration. 

We conducted the numerical calculation for an ultra-relativistic flow by setting $A=1.0\times 10^{12}$. 
The calculation continued till $\eta$ reached $\sim 10^6$, but it could not reach the stage of free expansion. 
We could not confirm by numerical calculations whether an ultra-relativistic flow leads to free expansion or not. 

On the other hand, when a flow is not so ultra-relativistic at the time of 
shock emergence, the acceleration comes to an end both in SS and in PP configuration. 
However there remains a difference between them; a spherical flow rather suddenly  finishes the acceleration while a PP flow approaches the end of acceleration asymptotically. 

\acknowledgments
This work has been partially supported by the grants in aid (16540213) of the Ministry of Education, Science, Culture, and Sports in Japan.

\end{document}